\documentclass[twocolumn,preprintnumbers,amsmath,amssymb,aps,prb,superscriptaddress,floatfix]{revtex4}
\usepackage{graphicx}
\usepackage{times}

\begin{document}

\title{Direct surface cyclotron resonance terahertz emission from a quantum cascade structure}

\author{Fran\c cois-R\'egis Jasnot}
\affiliation{Laboratoire Pierre Aigrain, Ecole Normale Sup\'erieure,
CNRS (UMR 8551), 24 rue Lhomond, 75231 Paris Cedex 05, France}

\author{Louis-Anne de Vaulchier}
\affiliation{Laboratoire Pierre Aigrain, Ecole Normale Sup\'erieure,
CNRS (UMR 8551), 24 rue Lhomond, 75231 Paris Cedex 05, France}

\author{Yves Guldner}
\affiliation{Laboratoire Pierre Aigrain, Ecole Normale Sup\'erieure,
CNRS (UMR 8551), 24 rue Lhomond, 75231 Paris Cedex 05, France}

\author{G\'erald Bastard}
\affiliation{Laboratoire Pierre Aigrain, Ecole Normale Sup\'erieure,
CNRS (UMR 8551), 24 rue Lhomond, 75231 Paris Cedex 05, France}

\author{Angela Vasanelli}
\affiliation{Laboratoire Mat\'eriaux et Ph\'enom\'enes Quantiques, Universit\'e Paris Diderot - Paris 7,
CNRS - UMR 7162, B\^atiment Condorcet, 75205 Paris Cedex 13, France}

\author{Christophe Manquest}
\affiliation{Laboratoire Mat\'eriaux et Ph\'enom\'enes Quantiques, Universit\'e Paris Diderot - Paris 7,
CNRS - UMR 7162, B\^atiment Condorcet, 75205 Paris Cedex 13, France}

\author{Carlo Sirtori}
\affiliation{Laboratoire Mat\'eriaux et Ph\'enom\'enes Quantiques, Universit\'e Paris Diderot - Paris 7,
CNRS - UMR 7162, B\^atiment Condorcet, 75205 Paris Cedex 13, France}

\author{Mattias Beck}
\affiliation{Institute of Quantum Electronics, ETH Z\"urich, CH-8093 Z\"urich, Switzerland}

\author{J\'er\^ome Faist}
\affiliation{Institute of Quantum Electronics, ETH Z\"urich, CH-8093 Z\"urich, Switzerland}




\begin{abstract}
A strong magnetic field applied along the growth direction of a semiconductor quantum well gives rise to a spectrum of discrete energy states, the Landau levels. By combining quantum engineering of a quantum cascade structure with a static magnetic field, we can selectively inject electrons into the excited Landau level of a quantum well and realize a tunable surface emitting device based on cyclotron emission. By applying the appropriate magnetic field between 0 and 12 T, we demonstrate emission from a single device over a wide range of frequencies (1-2 THz and 3-5 THz).
\end{abstract}

\maketitle


In 1960 well before the advent of diode lasers, it was proposed that semiconductors immerse in a static magnetic field were likely candidates for use as quantum amplifier and oscillators~\cite{Lax:1960}. A follow up on this idea was implemented in the 80's on a highly purified bulk germanium lightly doped with holes~\cite{Ivanov:1983, Unterrainer:1990}. In this system laser action in the terahertz (THz) region was obtained, at cryogenic temperatures, by cross combining intense magnetic and electric fields. More recently, semiconductor THz lasers have been demonstrated using quantum cascade laser concepts~\cite{Kohler:2002, Williams:2007}. Owing to band structure engineering and the use of guided optics a much higher degree of control on the spectral and spatial properties have been obtained in these last devices. Using quantum cascade technology distributed feedback lasers~\cite{Mahler:2004}, photonic crystal devices~\cite{Chassagneux:2009} and external cavities have been demonstrated\cite{Lee:2010,Qin:2009}. However, an inherent property of these lasers is that the spectral position of the optical gain is fixed, at a given current, and therefore to realise wavelength tunable devices it is necessary to employ mechanical systems.

In this paper we propose to combine the quantum engineering of a quantum cascade structure (QCS) with a static magnetic field to realise frequency tunable surface emitting electroluminescent devices. The specific strategy is the use of a resonant injector which allows to populate selectively a given Landau level (LL) of a quantum well. Besides, the fundamental concept of the QCS allows to magnify even further the inter-LL emission signal by increasing the number of periods, the extraction miniband of a period being the injection miniband of the next one. In our device electrons are injected into an excited state of a transition between two LLs arising from the extra confinement imposed by the magnetic field. It is in fact known that a magnetic field parallel to the growth axis of a quantum well structure breaks the two-dimensional parabolic energy dispersion of each subband $\varepsilon_i \left(\textbf{k}\right)$ into a ladder of discrete LLs with an energy $\varepsilon_{i,j} = E_i + \left(j+1/2\right) \hbar \omega_c$ and separated by the cyclotron energy $\hbar \omega_c = \hbar e B / m^{\ast}$, where $i$ is the subband index, $E_i$ the energy of the subband edge at zero magnetic field, $j$ the LL integer index, $B$ the magnetic field and $m^{\ast}$ the energy dependent electron effective mass. Experiments with magnetic field have been widely used to evaluate the different contributions of scattering mechanisms in complex QCSs~\cite{Ulrich:2000,Blaser:2002,Smirnov:2002,Scalari:2006,Scalari:2007,Leuliet:2006,Perelaperne:2007,Gomez:2008,Jasnot:2010}.

To achieve electrical injection into an excited LL we have conceived a QCS in which electrons are injected into the first excited subband $\left|2 \right\rangle$ of a large quantum well. Under the influence of a magnetic field the excited LLs arising from the first subband, $\left|1,j \right\rangle$, get in energy proximity with the first LL of the subband $\left|2,0 \right\rangle$. It has been theoretically and experimentally demonstrated that in the vicinity of crossing LLs  which belong to different subbands, a disorder activated transition becomes significant and allows a redistribution of the electron population between them~\cite{Regnault:2007}. By exploiting this mechanism electrons are therefore transferred, via an elastic scattering, from $\left|2,0 \right\rangle$ to $\left|1,j \right\rangle$ and subsequently generate photons by spontaneous emission between $\left|1,j \right\rangle$  and $\left|1,j-1 \right\rangle$. The optical dipole between LLs is perpendicular to the growth axis $Oz$ and thus photons can be emitted through the surface of the sample with a polarization $\varepsilon_x$ contrary to intersubband transition where light has a polarization $\varepsilon_z$ and is emitted through the side of the device.\cite{Williams:2007}

\begin{figure}
\centering
\includegraphics[width=0.4\textwidth]{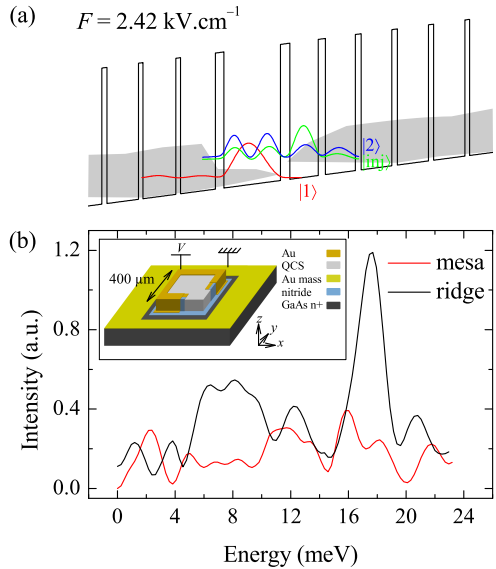}
\caption{(a) Schematic conduction band diagram and squared wave functions of one QCS period under an applied electric field. The radiative transition occurs between the first excited state $\left| 2 \right\rangle$ and the fundamental state $\left| 1 \right\rangle$ of the thick quantum well. Electrons are injected into $\left| 2 \right\rangle$ via the injector level $\left| \text{inj} \right\rangle$. The grey shaded regions correspond to the injection and extraction minibands. (b) Zero magnetic field electroluminescence spectra at a current density of $40\,\text{A.cm}^{-2}$ for the mesa (red) and the ridge (black) devices. The temperature is $4.5\,\text{K}$. Inset : schematic of the QCS with the specific processing to block intersubband emission with a polarization $\varepsilon_z$.}
\label{figure1}
\end{figure}

The sample used is a GaAs/Al$_{0.15}$Ga$_{0.75}$As THz quantum cascade structure, emitting at 4.3 THz. The active region is made of 80 identical periods, based on a main quantum well, where the radiative transition takes place between levels $\left|2\right\rangle$ and $\left|1\right\rangle$ ($E_{2 \to 1} \approx 18\,\text{meV} \to 4.3\,\text{THz}$), and an injection/extraction miniband (Figure~\ref{figure1}(a)). The detailed succession of wells and barriers (in Angstrom) for a single period is the following: \textbf{44} / 266 / \textbf{40} / 164 / \textbf{22} / 154 / \textbf{22} / 148 / \textbf{24} / 146 / \textbf{26} / 144 / \textbf{28} / 140 / \textbf{34} / 134 (Al$_{0.15}$Ga$_{0.75}$As barriers in bold). The $148\,\text{\AA}$ and $146\,\text{\AA}$ GaAs layers are Si-doped $2\times10^{16}\,\text{cm}^{-3}$. Two devices, specially conceived to observe surface emission, have been processed : a  mesa and a ridge. A schematic of the mesa is shown in the inset of Figure~\ref{figure1}(b). The top of the mesa is a  $300\,\mu\text{m} \times 300\,\mu$m  GaAs free surface surrounded by an edge of gold of $50~\mu$m width to realise electrical connection. The gold layer is separated from the mesa by a $0.5\,\mu$m layer of silicon nitride preventing electrical short-circuit. As all the sides of the device are covered by gold, only surface emission is possible from the device. A $300\,\mu$m wide and $3.6\,\text{mm}$ long ridge was also processed. The top surface of this device consists of two 50~$\mu$m wide bands of gold separated by a $200\,\mu$m wide GaAs free surface. This geometry allows the observation of a luminescence signal from both the surface and the facets.

Electroluminescence spectra of the two devices at zero magnetic field are shown in Figure~\ref{figure1}(b). The spectra are measured for the same current density with a Bruker Vertex 80V FTIR in the step scan mode on a liquid-helium cooled Si-bolometer, the signal being detected with a lock-in amplifier. In our setup, the detector is turned toward the surface of the sample. The spectrum measured from the ridge consists of a narrow intersubband electroluminescence peak centered at an energy of $\sim 18\,\text{meV}$ in agreement with the calculated energy $E_{2 \to 1}$. This indicates that, although the detector is turned towards the surface of the sample, the $\varepsilon_z$ polarized light emitted through the facet is nonetheless detected. This is due to the diffraction of the light on the side of the ridge and to its multi-reflection on the tube around the sample. In addition, a weak broad peak is highlighted between 4 and $14\,\text{meV}$ and is attributed to the black body of the sample. By using the Wien law, we find a temperature of $\sim 17\,\text{K}$. On the contrary, no peak is visible on the mesa spectrum. No black body radiation is observed either because the current flowing through the device is lower. The size of the sample is indeed reduced by a factor of 7. These two spectra demonstrate the efficiency of our mesa processing to block the $\varepsilon_z$ polarized emission.

\begin{figure}
\centering
\includegraphics[width=0.35\textwidth]{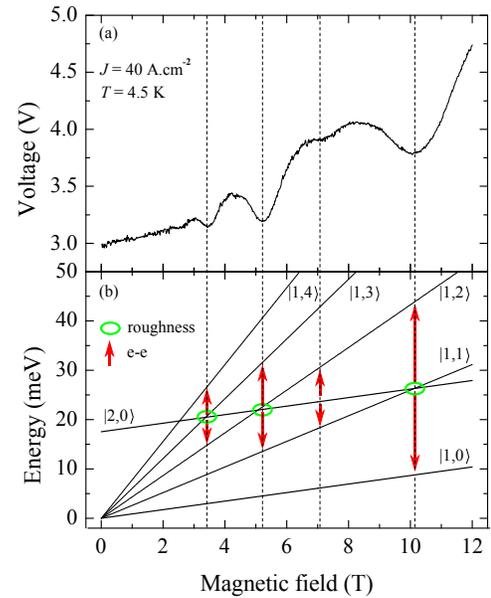}
\caption{(a) Voltage as a function of the magnetic field at constant current density  for the mesa device. (b) Plot of the LLs $\left| 2,0\right\rangle$  and $\left| 1, j \right\rangle$. Two elastic scattering mechanisms are displayed : electron-electron interaction (red arrows) and interface roughness (green circles).}
\label{figure2}
\end{figure}

\begin{figure}
\centering
\includegraphics[width=0.5\textwidth]{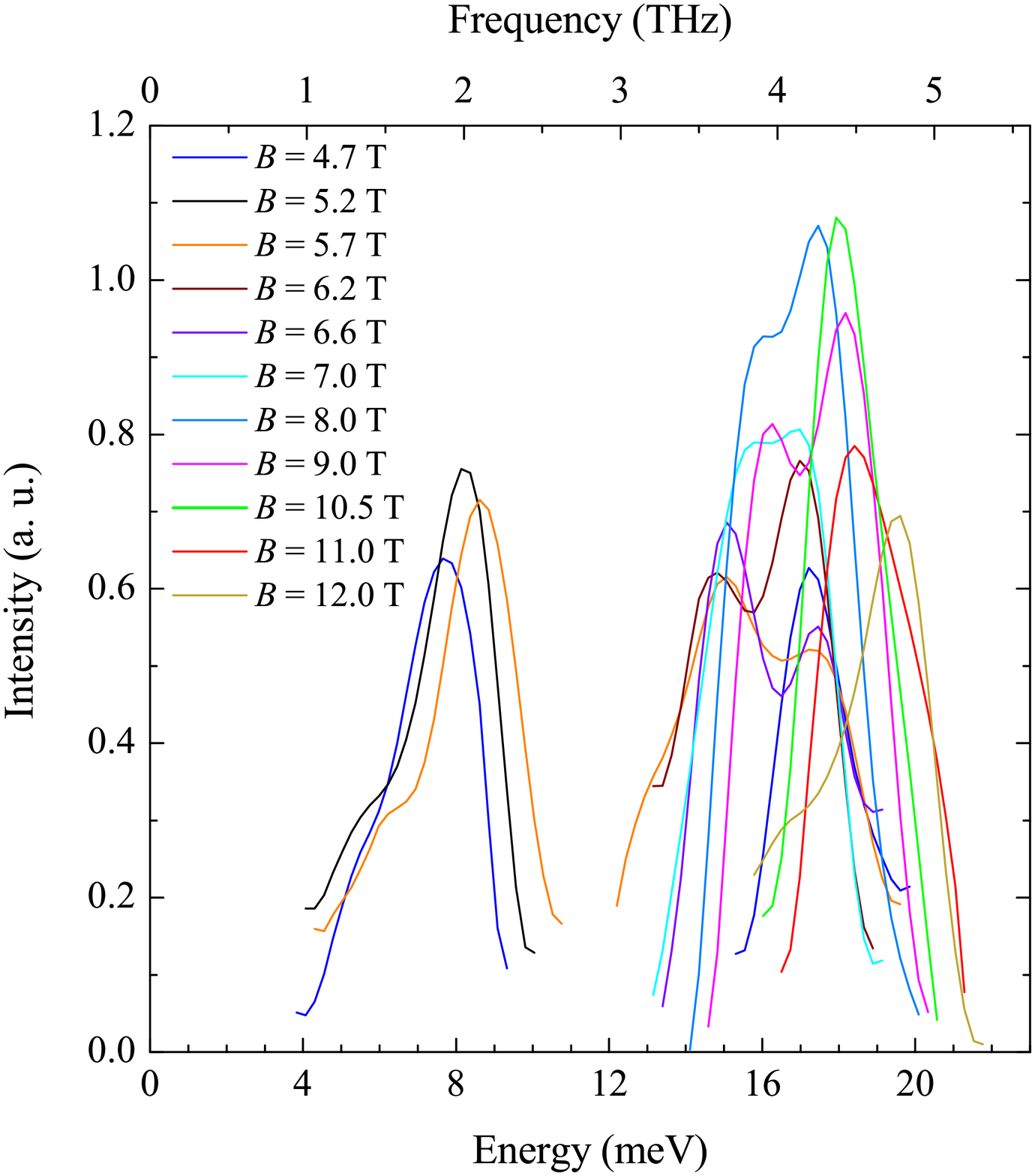}
\caption{Emission spectra for different magnetic field from $B=4.7$ to $12.0\,\text{T}$ at fixed current density $J = 40\,\text{A} . \text{cm}^{-2}$. The resolution is $1.2\,\text{meV}$ and the temperature is $4.5\,\text{K}$.}
\label{figure3}
\end{figure}

The mesa device was mounted inside an insert at the center of a superconducting coil capable of producing fields up to $12\,\text{T}$ such that the magnetic field lines are perpendicular to the plane of the quantum wells.  Figure~\ref{figure2}(a) shows the evolution of the voltage as a function of the magnetic field at fixed current density. The voltage exhibits oscillations as a function of $B$ superimposed on a continuous increasing background which is attributed to the magnetoresistance of the contact~\cite{Leuliet:2006,Gomez:2008}. Three main minima are located at $B=3.5\,\text{T}$, $5.2\,\text{T}$ and $10.5\,\text{T}$. An extra minimum, less pronounced, is present at $B=7.2\,\text{T}$. These features are a consequence of the electron lifetime modulation on the level $\left|2,0\right\rangle$ of the thick quantum well. Figure~\ref{figure2}(b) shows the LLs $\left|2,0\right\rangle$ and $\left|1,j\right\rangle$ as a function of the magnetic field.  Dashed vertical lines are used to indicate the values of the magnetic field for which  $\left|2,0\right\rangle$ is in resonance with one of the $\left|1,j\right\rangle$ and/or halfway between two LLs of $\left|1\right\rangle$.

 Two elastic mechanisms can occur in such situation: interface roughness or electron-electron (Auger like) scattering. The two mechanisms can occur at $B=3.5, 5.2$ and $10.5\,\text{T}$  whereas at $B=7.2\,\text{T}$ only electron-electron scattering can occur. Interface roughness has been widely reported as the main mechanism in GaAs/AlGaAs THz QCS~\cite{Scalari:2007,Perelaperne:2007}. Signature of electron-electron scattering has been also reported in transport\cite{Kempa:2002,Kempa:2003} and laser power measurements\cite{Scalari:2004}.

\begin{figure}
\centering
\includegraphics[width=0.5\textwidth]{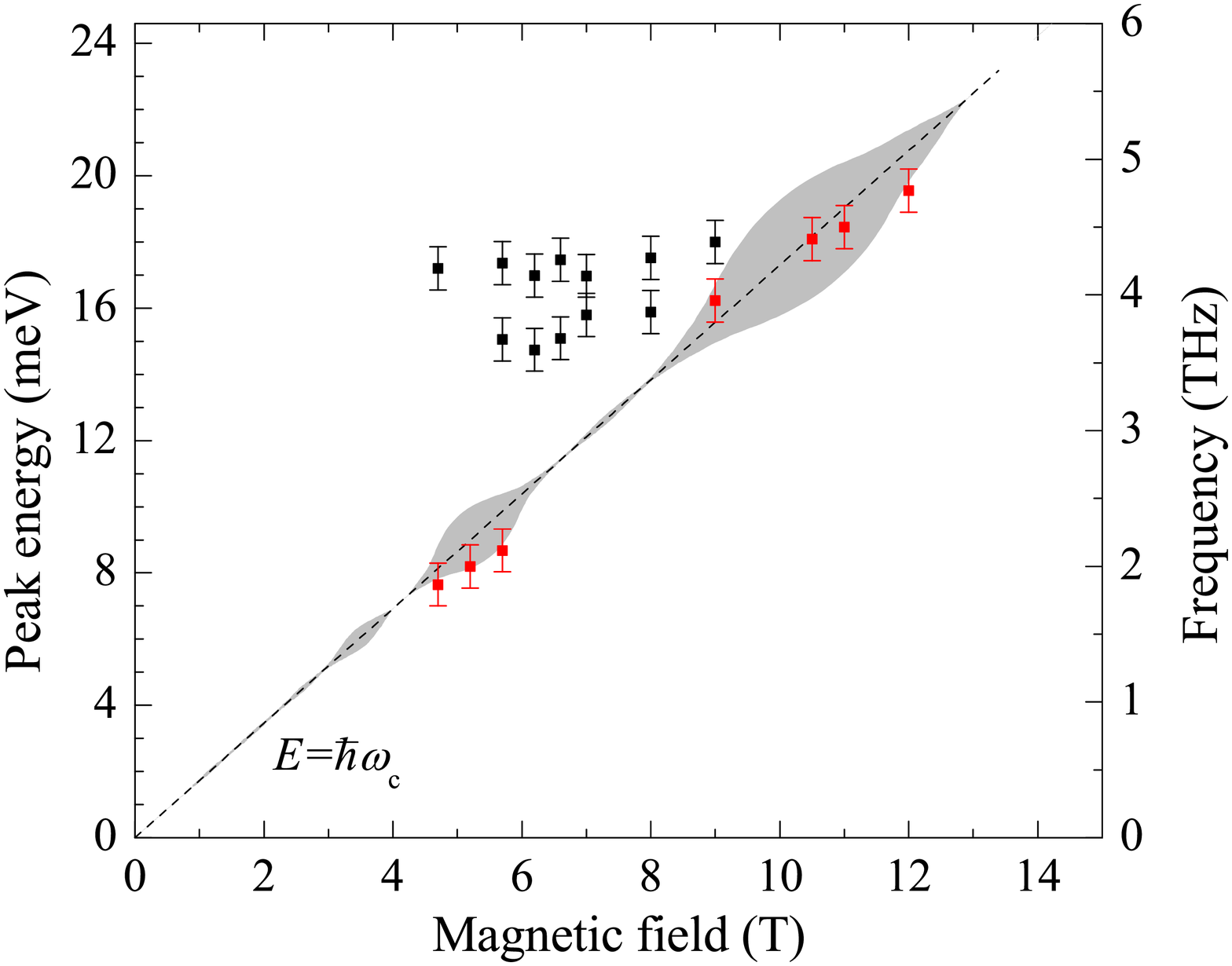}
\caption{Emission peak energy at different magnetic field and fixed current density. Error bars are due to the experimental resolution of our spectra ($1.2\,\text{meV}$).  Red squares correspond to the cyclotron emission peaks. The black squares correspond to disorder activated forbidden transitions. The grey areas are proportional to the number of injected electrons in the LLs as deduced from the experimental transport data of Figure~\ref{figure2}(a).}
\label{figure4}
\end{figure}

Experimental spectra of the mesa at different magnetic fields are presented in Figure~\ref{figure3}.  For the sake of clarity the residual background and the black body appearing at high magnetic field have been removed. Two main series of peaks can be observed, one around $8\,\text{meV}$ and another one around $18\,\text{meV}$. These two characteristic energies correspond to the cyclotron energies in the vicinity of 5.2 and $10.5\,\text{T}$ respectively. It has to be noted that around 6 and $12\,\text{meV}$, which corresponds to 3.5 and $7.2\,\text{T}$ respectively, no such series could be measured. This is related to the weak features of the transport measurement minima at these magnetic fields as reported in Figure~\ref{figure2}(a). It can also be seen that several spectra show two or three peaks rather than one only. In order to analyse the data we have represented in Figure~\ref{figure4} the peak energy of each emission spectra in Figure~\ref{figure3} as a function of magnetic field. The data reported in red squares correspond to the cyclotron emission and follow the $\hbar \omega_c$ law  ($m^{\ast} = 0.067 m_0$ for GaAs) indicated in dashed line.  As expected red points only show up in the grey areas where LLs are populated with electrons (see caption).

In addition to the cyclotron emission peaks, several others are measured. They are reported in black squares and two series can be identified; one at a roughly constant energy of $18\,\text{meV}$ between 5 and $9\,\text{T}$, and another one, in the similar $B$ range, $2\,\text{meV}$ below. These two characteristic energies are close to the intersubband transitions $E_{2 \to 1}$ and $E_{\text{inj} \to 1} $ respectively. As shown in the first part of this paper,  light emission with a polarization $\varepsilon_z$ is blocked in our device. Therefore, even if these peaks correspond to the intersubband energy, they cannot originate from $\vert 2,0 \rangle \rightarrow \vert 1,0 \rangle$ emission, that is polarized perpendicular to the surface. It could originate from carrier localization due to disorder\cite{Scalari:2004}. Also, it is worth noting that theoretical considerations have demonstrated that normally forbidden transitions in ideal structures could be activated by disorder. Specifically, Regnault \textit{et al.} have shown that the disorder induced coupling between LLs could be responsible for transitions in the wrong polarization. This mechanism could explain the two series at 18 and $16\,\text{meV}$ as electron-electron interaction can couple LLs. This observation implies that the widely used picture of energy-degenerate LLs, crossing without interaction has to be revisited.

In conclusion we report a magnetic field tunable surface emitting device in the $1-2$ and $3-5\,\text{THz}$ range based on cyclotron emission. The study of the CR emission allows the identification of the dominant scattering mechanisms occurring in the QCS. Our experiments show that QCS are good candidates for realizing Landau-level laser~\cite{Aoki:1986} or tunable THz VCSEL. In order to obtain a population inversion between LLs $\left|j\right\rangle$ and $\left|j+1\right\rangle$, non equidistant LLs are necessary in order to dissymetrise the absorption and emission transition energies between LLs. This situation could occur in our sample by taking advantage of the disorder inducing an inhomogeneous broadening of the LLs at large magnetic field. Additionally, this situation could occur in QCS based on low gap materials, such as InAs, in which the non-parabolicity could be exploited.


\end{document}